\documentclass[12pt,letterpaper,oneside]{article}
\usepackage{amssymb,tabularx,multirow,amsmath,natbib, epsfig, threeparttable, amstext, subfigure,xcolor,bbm}
\usepackage{pdflscape}
\usepackage{afterpage}
\usepackage{capt-of}
\usepackage{rotating}
\usepackage[makeroom]{cancel}
\usepackage{bbm}
\usepackage{dsfont}
\usepackage{comment}

\def\spacingset#1{\renewcommand{\baselinestretch}%
{#1}\small\normalsize} \spacingset{1}

\usepackage[dvips,top=1.2in,bottom=0.65in,left=1.15in,right=1.15in,includefoot]{geometry}

\usepackage[margin=20pt,labelfont={bf,small},textfont={it,small},indention=0.5cm]{caption}

\begin{document}
  \title{\bf Geostatistical Capture-Recapture Models}
  \author{Mevin B. Hooten$^1$, Michael R. Schwob$^1$, Devin S. Johnson$^2$, \\ and Jacob S. Ivan$^3$ \\ \vspace{.02in} \\
    $^1$ Department of Statistics and Data Sciences \\ The University of Texas at Austin \\ \vspace{.02in} \\ $^2$ Pacific Islands Fisheries Science Center \\ National Marine Fisheries Service \\ \vspace{.02in} \\ $^3$ Colorado Parks and Wildlife}
  \maketitle

\begin{abstract}
Methods for population estimation and inference have evolved over the past decade to allow for the incorporation of spatial information when using capture-recapture study designs.  Traditional approaches to specifying spatial capture-recapture (SCR) models often rely on an individual-based detection function that decays as a detection location is farther from an individual's activity center.  Traditional SCR models are intuitive because they incorporate mechanisms of animal space use based on their assumptions about activity centers.  We modify the SCR model to accommodate a wide range of space use patterns, including for those individuals that may exhibit traditional elliptical utilization distributions.  Our approach uses underlying Gaussian processes to characterize the space use of individuals.  This allows us to account for multimodal and other complex space use patterns that may arise due to movement.  We refer to this class of models as geostatistical capture-recapture (GCR) models.  We adapt a recursive computing strategy to fit GCR models to data in stages, some of which can be parallelized.  This technique facilitates implementation and leverages modern multicore and distributed computing environments.  We demonstrate the application of GCR models by analyzing both simulated data and a data set involving capture histories of snowshoe hares in central Colorado, USA.   
\end{abstract}

\noindent%

\spacingset{1.45} 
\section{Introduction}
Bayesian models for spatial capture-recapture (SCR) data have become popular for estimating wildlife population demographics \citep{royle2013spatial}.  Conventional SCR models utilize data that comprise detections of a subset of individual animals from a wildlife population at an array of detectors (often referred to as ``traps'').  By accounting for spatial structure in the underlying movement process of the individuals, SCR models can be used to infer population abundance and density (in addition to other quantities in generalized SCR models; \citealt{tourani2022review}).  We present a new model formulation that relaxes the conventional assumption of latent activity centers while still accounting for spatially structured species distributions and space use patterns.  Our model links individual-level detection probability to latent continuous spatial random fields and thus we refer to it as a geostatistical capture-recapture (GCR) model.  The GCR model is also amenable to multi-stage computing strategies that leverage parallel computing environments to accelerate and stabilize the implementation.   

In what follows, we present the conventional SCR model formulation to introduce our statistical notation and then highlight the components we modify in the subsequent methods section.  We follow the parameter-expanded data augmentation (PX-DA) approach of \citet{royle2012parameter} in the SCR framework \citep{royle2008hierarchical, royleyoung2008hierarchical}.  For a set of traps that detect individuals at locations $\mathbf{x}_l$ for $l=1,\ldots,L$ during sampling periods (or occasions) $j=1,\ldots,J$, we observe binary detection/nondetection measurements for a set of $n$ observed individuals $y_{i,l,j}$ where $i$ denotes individual.  It is often assumed that $y_{i,l,j}$ are independent conditioned on individual $i$ and trap $l$.  Thus, the count $y_{i,l}=\sum_{j=1}^J y_{i,l,j}$ represents the number of detections of individual $i$ at trap $l$ with conditional mixture binomial distribution 
\begin{equation}
  y_{i,l} \sim 
  \begin{cases}
    \text{Binom}(J,p_{i,l}) &\mbox{, } z_i = 1 \\   
    0 &\mbox{, } z_i = 0 \\   
  \end{cases} \;, 
  \label{eq:scr_datamodel}
\end{equation}
\noindent where $z_i$ is a latent population membership indicator for $i=1,\ldots,M$, with $M$ chosen so that it provides a realistic upper bound for population abundance (usually $M>>n$).  In this type of PX-DA scenario, the data are augmented with all-zero capture histories such that $y_{i,l}=0$ for all $i>n$ and the latent indicators are modeled as $z_i\sim \text{Bern}(\psi)$ \citep{royle2009analysis}.   

Most formulations of SCR models account for heterogeneity in detection probability $p_{i,l}$ by relating it to the distance between an unobserved individual activity center $\mathbf{c}_i$ and the trap location $\mathbf{x}_l$ such that
\begin{equation}
  g(p_{i,l})=\alpha+\beta\cdot||\mathbf{c}_i-\mathbf{x}_l||^2 \;.
  \label{eq:scr_link}
\end{equation}
The term $\alpha$ accounts for baseline detection probability and $\beta$ (often expressed as $1/2\sigma^2$) controls the rate of decay in detection probability as the distance between activity center and trap location increases.  The functional form of (\ref{eq:scr_link}) implies a radial space use pattern by all individuals, and this aspect of conventional SCR models has been debated \citep[e.g.,][]{ivan2013using,royle2013integrating,fuller2016estimating, efford2019non, sutherland2019oscr}. Furthermore, the link function $g$ can be specified based on the study design or chosen based on convenience.  For example, \citet{royle2008hierarchical} described how the `cloglog' link function may accommodate data that arise as a censoring of counts based on multiple detections during a single time period.  

Our reformulation of the SCR model uses a geostatistical representation of the detection function.  The resulting GCR model is more robust to departures from individual elliptical space use patterns.  For example, \citet{wilson2010accounting} found that bobcat (\emph{Lynx rufus}) space use distributions could contain three distinct core areas of higher intensity use.  Similarly, \citet{kordosky2021landscape} observed some individual fishers (\emph{Pekania pennanti}) with multiple core areas.  Some bird species construct multiple nests during breeding season which may lead to complex space use patterns \citep{macqueen2023adaptive}.  Our GCR model is flexible enough to accommodate these types of species and individuals with irregular space use distributions.     

\section{Methods}
\subsection{Geostatistical Capture-Recapture (GCR) Model}
To generalize the SCR model, we retain the data model in (\ref{eq:scr_datamodel}) and develop a robust stochastic process model for the detection probabilities $p_{i,l}$.  A common criticism of conventional SCR models is that the link function in (\ref{eq:scr_link}) implies a homogeneous radial space use distribution for each individual $i$ that may not be realistic.  Extensions that allow for heterogeneity in the pattern of activity centers $\mathbf{c}_i$ using inhomogeneous point process models have been proposed \citep[e.g.,][]{royle2013integrating,royle2018unifying}, but many studies assume the activity centers are independent and complete spatial random such that $\mathbf{c}_i\sim \text{Unif}({\cal S})$ for all $i$ (for compact study region ${\cal S}$).  

Other approaches have generalized the detection function; for example, \citet{fuller2016estimating} incorporated least-cost distance to landscape features.  Their approach was designed for cases when environmental characteristics that affect movement are known (e.g., river corridors; \citealt{sutherland2015modelling}), but may not be helpful when barriers or movement corridors are unobserved.  \citet{royle2013integrating} developed a hybrid resource selection function (RSF) SCR model that allowed for heterogeneity in individual space use and in the broader species distribution.  The hybrid RSF-SCR approach allows the individual locations to vary according to spatially referenced covariates and a general attraction to a single activity center.  Similarly, a variety of new approaches to explicitly accommodate fine scale animal movement in SCR models \citep[e.g.,][]{mcclintock2022integrated} have been proposed, but may require additional auxiliary data sources.     

Many of the aforementioned extensions to SCR models could be generalized to incorporate additional mechanisms associated with the processes of animal movement and space use (e.g., multiple activity centers, barriers and corridors to movement, etc.), but as the models become more complex they require more parameters and hence more data to learn those parameters effectively.  To increase the set of options for modeling SCR data when they are limited and$/$or the mechanisms are less well known, we offer a flexible and parsimonious stochastic model for detection probability that accommodates spatial structure and irregularity in the species distribution and space use pattern together.    

For a position $\mathbf{x}$ of individual $i$, we consider the detection function $p_i(\mathbf{x})$ (for $0<p_i<1$) as a smooth stochastic process in space.  If the function $p_i(\mathbf{x})$ is proportional to the probability of an individual choosing to occur in that region of space then it could be referred to as a resource selection probability function (RSPF) \citep{lele2009new, hooten2020animal} and admits the notion of animal movement as an underlying mechanism that leads to detection in wildlife surveys \citep{hooten2017animal}.       

We use a Gaussian process representation to construct a model for $p_i(\mathbf{x})$ such that it exists for any position $\mathbf{x}$.  First, without loss of generality, we consider a latent Gaussian process $v_i(\mathbf{x})$ with mean $\mu$ and covariance function between any two locations $l$ and $\tilde{l}$, defined as   
\begin{equation}
  \text{cov}(v_i(\mathbf{x}_l),v_i(\mathbf{x}_{\tilde{l}}))=\sigma^2 R_{l,\tilde{l}}=\sigma^2\exp\left(-\frac{(\mathbf{x}_l-\mathbf{x}_{\tilde{l}})'(\mathbf{x}_l-\mathbf{x}_{\tilde{l}})}{\theta^2}\right) \;, 
  \label{eq:gcr_cov}
\end{equation}
\noindent where this form can be generalized to accommodate any valid covariance function (e.g., the Matern class; \citealt{stein2012interpolation}).  We connect the Gaussian process to the detection probability by $g(p_i(\mathbf{x}_l))=v_i(\mathbf{x}_l)$ for a valid link function $g(\cdot)$ (e.g., logit, probit, etc).  By combining the SCR data model in (\ref{eq:scr_datamodel}) with a latent Gaussian process, we have a hierarchical zero-inflated binomial model with $g(p_i(\mathbf{x}_l))=v_i(\mathbf{x}_l)$.  

To complete the Bayesian model specification, only the spatial range (i.e., lengthscale) parameter $\theta$, mean $\mu$, and variance $\sigma^2$ are unknown random variables in the model.  We specify a normal prior for $\mu$ such that $\mu\sim \text{N}(\mu_0,\sigma^2_0)$, where $\mu_0=0$ can often be assumed. When a probit link is selected for $g$, we set $\sigma^2=1$, but an inverse gamma prior for $\sigma^2$ is conjugate if desired.  A variety of options for prior distributions are available for $\theta$, however, one particularly useful prior from a computational perspective is the discrete uniform such that $\theta \sim \text{DiscUnif}(\Theta)$ on the finite discrete support set $\Theta$ \citep{diggle2002bayesian}.  By selecting a grid of values for $\Theta$ that span a range of realistic scales for the spatial structure in $v_i$ (and hence $p_i$), we can precompute and store the matrix quantities that are necessary to implement the model.  This can be done in advance of model fitting and in parallel.    

\subsection{Recursive Bayesian Implementation}
A traditional MCMC algorithm can be used to fit either the original SCR or our new GCR model, but such algorithms often suffer from poor mixing due to the need to tune the $\mathbf{c}_i$ updates in the case of (\ref{eq:scr_link}) or all $p_{i,l}$ updates in the case of our GCR model.  \cite{hooten2023multistagecr} described an approach to fit CR models using multistage algorithms that leverage parallel computing resources.  They also demonstrated that helpful adjustments to the model components arise when considering recursive implementations.  We show that similar strategies can be applied to fit GCR models to SCR data.   

We seek a recursive Bayesian procedure that involves computational stages to fit the GCR model to data.  This approach was inspired by the meta-analytic two-stage MCMC procedure described by \citet{lunn2013fully} and generalized by \citet{hooten2021making} and \citet{mccaslin2021tarb} where it was referred to as ``prior-proposal recursive Bayesian'' (PPRB) computation.  In general, PPRB is a simple approach to obtain posterior inference based on a procedure with parallelizable components, but \cite{hooten2023multistagecr} showed that it can be particularly useful for fitting CR models with heterogeneous detection probabilities.  

The full Bayesian GCR model we presented in Section 2.1 results in the posterior distribution 
\begin{equation}
  [\mathbf{V},\mathbf{z},\psi,\mu,\sigma^2,\theta|\mathbf{Y}]\propto \left(\prod_{i=1}^M \left(\prod_{l=1}^L [y_{i,l}|p_{i,l},z_i]\right) [z_i|\psi][\mathbf{v}_i|\mu,\sigma^2,\theta]\right)[\psi][\mu][\sigma^2][\theta]\;, 
  \label{eq:big_post}
\end{equation}
\noindent for $\mathbf{V}\equiv(\mathbf{v}_1,\ldots,\mathbf{v}_M)$, where $[\mathbf{v}_i|\mu,\sigma^2,\theta]$ represents the latent Gaussian process that is linked to $\mathbf{p}_i$, and $\mathbf{P}\equiv(\mathbf{p}_1,\ldots,\mathbf{p}_M)$.  We specify the prior for $\psi$ as $[\psi]=\text{Beta}(\alpha_\psi,\beta_\psi)$ with hyperparameters $\alpha_\psi$ and $\beta_\psi$ either set based on prior information or to approximate a scale prior \citep{link2013cautionary}.  A conventional implementation of this model using MCMC would require $M\cdot L$ Metropolis-Hastings updates, each of which could potentially involve proposal distributions that require tuning due to the lack of conjugacy when updating $p_{i,l}$.  Thus, although the GCR model is more flexible than the classical SCR model, we would expect similar instability issues with the associated MCMC algorithm.  Approaches based on integrated likelihoods can sometimes be helpful \citep[e.g.,][]{borchers2008spatially, efford2009density, efford2011estimation, king2016capture} and numerical integration approaches commonly used to fit SCR models are usually more stable, but also computationally intensive.

To remedy the aforementioned issues, we fit the GCR  model following the capture-recapture recursive Bayesian implementation proposed by \citet{hooten2023multistagecr} which is comprised of two main stages.  In the first stage, we condition on the observed individual capture histories $\mathbf{y}_i$ for $i=1,\ldots,n$ and marginalize over the latent process $\mathbf{v}_i$.  In the second stage, we subsample from the first stage output using the conditional distribution for $n$.  This procedure allows us to obtain a final MCMC sample from the posterior distribution associated with our GCR model:   
\begin{equation}
  [\mu,\sigma^2,\theta,\psi|\mathbf{y}_1,\ldots,\mathbf{y}_n, n] \propto \left(\prod_{i=1}^n \left[ \mathbf{y}_i | \mu, \sigma^2, \theta , \mathbf{y}'_i\mathbf{1}>0\right] \right)[n|\mu, \sigma^2, \theta, \psi][\mu][\sigma^2][\theta][\psi] \;, 
  \label{eq:full_post}
\end{equation}
\noindent with component distributions as described in what follows.  We show how to derive (\ref{eq:full_post}) in Appendix A.   

For the first stage, we condition on the knowledge that the observed individuals were captured at least once which yields the integrated data model 
\begin{equation}
   \left[ \mathbf{y}_i | \mu, \sigma^2, \theta , \mathbf{y}'_i\mathbf{1}>0\right]= \int [\mathbf{y}_i|\mathbf{p}_i,\mathbf{y}'_i\mathbf{1}>0][\mathbf{v}_i|\mu,\sigma^2,\theta,\mathbf{y}'_i\mathbf{1}>0] d \mathbf{v}_i \;, \label{eq:int_data_mod}
\end{equation}
\noindent where the conditional data model for $\mathbf{y}_i$ given at least one detection is
\begin{align}  
  [\mathbf{y}_i|\mathbf{p}_i,\mathbf{y}'_i\mathbf{1}>0]&=\frac{\text{Pr}(\mathbf{y}'_i\mathbf{1}>0|\mathbf{y}_i,\mathbf{p}_i)[\mathbf{y}_i|\mathbf{p}_i]}{\sum_{\mathbf{y}}\text{Pr}(\mathbf{y}'\mathbf{1}>0|\mathbf{y},\mathbf{p}_i)[\mathbf{y}|\mathbf{p}_i]} \;, \\  
  &=\frac{\mathds{1}_{\{\mathbf{y}'_i\mathbf{1}>0\}}[\mathbf{y}_i|\mathbf{p}_i]}{\sum_{\mathbf{y}}\mathds{1}_{\{\mathbf{y}'\mathbf{1}>0\}}[\mathbf{y}|\mathbf{p}_i]} \;, \\
  &=\frac{\mathds{1}_{\{\mathbf{y}'_i\mathbf{1}>0\}}[\mathbf{y}_i|\mathbf{p}_i]}{1-\prod_{l=1}^L (1-p_{i,l})^J} \;. \label{eq:conddatamod}
\end{align}
\noindent The denominator in (\ref{eq:conddatamod}) is equal to $1-[\mathbf{y}=\mathbf{0}|\mathbf{p}_i]$ for the binomial model $[\mathbf{y}_i|\mathbf{p}_i]=\prod_{l=1}^L [y_{i,l}|p_{i,l}]$, such that $[y_{i,j}|p_{i,l}]\equiv \text{Binom}(J,p_{i,l})$.  

The process model for $\mathbf{v}_i$ conditioned on at least one detection for individual $i$ is  
\begin{align}
   [\mathbf{v}_i|\mu,\sigma^2,\theta,\mathbf{y}'_i\mathbf{1}>0] &= \frac{\text{Pr}(\mathbf{y}'_i\mathbf{1}>0|\mathbf{p}_i,\mu,\sigma^2,\theta)[\mathbf{v}_i|\mu,\sigma^2,\theta]}{\int \text{Pr}(\mathbf{y}'_i\mathbf{1}>0|\mathbf{p},\mu,\sigma^2,\theta)[\mathbf{v}|\mu,\sigma^2,\theta]d\mathbf{v}}  \;, \\
   &=\frac{(1-\prod_{l=1}^L (1-p_{i,l})^J)[\mathbf{v}_i|\mu,\sigma^2,\theta]}{\int (1-\prod_{l=1}^L (1-p_{l})^J)[\mathbf{v}|\mu,\sigma^2,\theta]d\mathbf{v}} \;,
\end{align}
\noindent which together with the conditional data model implies the integrated data model in  (\ref{eq:int_data_mod}) can be expressed as 
\begin{equation}
   \left[ \mathbf{y}_i | \mu, \sigma^2, \theta , \mathbf{y}'_i\mathbf{1}>0\right]=\frac{\int \mathds{1}_{\{\mathbf{y}'_i\mathbf{1}>0\}}[\mathbf{y}_i|\mathbf{p}_i][\mathbf{v}_i|\mu,\sigma^2,\theta]d\mathbf{v}_i}{\int (1-\prod_{l=1}^L (1-p_{l})^J)[\mathbf{v}|\mu,\sigma^2,\theta]d\mathbf{v}} \;. \label{eq:int_data_mod2}
\end{equation}

Thus, in the first computing stage, we draw a sample for $\mu$, $\sigma^2$, $\theta$, and $\psi$, from the temporary posterior which is proportional to  
\begin{equation}
  \left(\prod_{i=1}^n \left[ \mathbf{y}_i | \mu, \sigma^2, \theta , \mathbf{y}'_i\mathbf{1}>0\right] \right)[\mu][\sigma^2][\theta][\psi] \;, 
\end{equation}
while noting that $\psi$ does not appear in the integrated data model and thus a Monte Carlo (MC) sample can be drawn from its prior $[\psi]$ directly. 

In the second computing stage, we sample from the full posterior distribution in (\ref{eq:full_post}) using the first stage temporary posterior as a proposal for $\mu$, $\sigma^2$, $\theta$, and $\psi$.  This results in a block update for all parameters using the Metropolis-Hastings ratio 
\begin{equation}
  \frac{[n|\mu^{(*)}, \sigma^{2(*)}, \theta^{(*)}, \psi^{(*)}]}{[n|\mu^{(k-1)}, \sigma^{2(k-1)}, \theta^{(k-1)}, \psi^{(k-1)}]} \;,
  \label{eq:stage2n}
\end{equation}
\noindent for MCMC iterations $k=1,\ldots,K$ and where the $(*)$ indicates a random draw from the first stage sample with replacement.

The conditional distribution for $n$ in (\ref{eq:full_post}) and (\ref{eq:stage2n}) is 
\begin{equation}
  [n|\mu, \sigma^2, \theta, \psi]=\int [n| \mathbf{P}, \psi][\mathbf{V}|\mu,\sigma^2,\theta]d\mathbf{V}\;,
  \label{eq:n_cond}
\end{equation}
\noindent where $[\mathbf{V}|\mu,\sigma^2,\theta]=\prod_{i=1}^M [\mathbf{v}_i|\mu,\sigma^2,\theta]$ is a product of multivariate Gaussian densities. The crux of this second computing stage lies in the evaluation of (\ref{eq:n_cond}). 
 We need to approximate the integral, either through numerical quadrature or MC integration by sampling from the multivariate Gaussian density.  However, the conditional distribution in (\ref{eq:n_cond}) can be computed in parallel and stored for recall as necessary in the first and second stage of the procedure.  Also, \cite{hooten2023multistagecr} showed that $[n| \mathbf{P}, \psi]$ could be specified as Poisson with intensity $\sum_{i=1}^M \psi(1-\prod_{l=1}^L(1-p_{i,l})^J)$.  This Poisson model for $n$ is much more computationally efficient to evaluate than the Poisson-binomial model that is implied in the original SCR model.  The Poisson model specification for $[n| \mathbf{P}, \psi]$ also yields nearly identical results as $M\rightarrow\infty$, thus we retain it in our implementation of the model that follows.    

\subsection{Inference and Prediction}
In terms of statistical inference, the GCR model allows us to quantify posterior characteristics of the parameters $\mu$, $\sigma^2$, $\theta$, and $\psi$.  However, most wildlife biologists are interested in inference associated with abundance $N$, the total number of individuals in the study population.  Following \cite{hooten2023multistagecr}, we sample the number of undetected individuals in our study population as $N_0^{(k)} \sim \text{Pois}(\bar{\psi}^{(k)}(M-n))$ and let $N^{(k)}=n+N_0^{(k)}$ (for $k=1,\ldots,K$) as a posterior realization of abundance.  The Poisson rate parameter can be computed as       
\begin{equation}
  \bar{\psi}^{(k)}=\int \left(\frac{\psi^{(k)}\prod_{l=1}^L(1-p_l)^J}{\psi^{(k)}\prod_{l=1}^L(1-p_l)^J+1-\psi^{(k)}}\right) [\mathbf{v}|\mu^{(k)},\sigma^{2(k)},\theta^{(k)}]d\mathbf{v} \;.
  \label{eq:psibar}
\end{equation}
The parenthetical term in (\ref{eq:psibar}) is the full-conditional probability of membership for an undetected individual in our population.  There are $M-n$ potential undetected individuals, thus we have the Poisson intensity $\bar{\psi}(M-n)$.  Posterior mean abundance can be computed using Monte Carlo integration as:  $\text{E}(N|\mathbf{Y})\approx \sum_{k=1}^K N^{(k)}/K$.   

This procedure for inferring abundance can be interpreted similarly to the non-spatial CR setting.  Our estimate of $N$ represents the capturable individuals in our population under study; that is, the population susceptible to our trap array during the study. 
 Our geostatistical model for detection probability accounts for individuals that may spend time outside the trap array, but unlike most other SCR models, it does not depend on latent activity centers explicitly (nor the associated assumptions about circular space use patterns).      

\cite{royle2008hierarchical} presented an excellent review of approaches for estimating effective sample area and formulations of density (i.e., abundance per unit area).  Similar approaches could be developed for the GCR model setting in cases where density, rather than abundance is of interest.  We focus on space use patterns in what follows and the ability of the GCR to account for multimodel and asymmetric characteristics of space use that may arise in different ways.  Thus, we estimate trap-specific detection probability $p_{i,l}$ for all individuals in the population and infer the individual-level space use patterns (i.e., the utilization distributions; \citealt{worton1989kernel}) using the posterior predictive distribution of $\tilde{\mathbf{v}}_i$.  We first obtain an MCMC sample of $\mathbf{v}_i$ from its full-conditional distribution based on the original joint distribution in (\ref{eq:big_post}), then sample $\tilde{\mathbf{v}}_i$ from its predictive full-conditional distribution given $\mathbf{v}_i$ by composition sampling with kriging (Appendix B).  The full-conditional distribution of $\mathbf{v}_i$ is
\begin{equation}
  [\mathbf{v}_i|\cdot]\propto[\mathbf{y}_i|\mathbf{p}_i][\mathbf{v}_i|\mu,\sigma^2,\theta] \;, 
  \label{eq:vfullcond}
\end{equation}
\noindent where $[\mathbf{y}_i|\mathbf{p}_i]=\prod_{l=1}^L[y_{i,l}|p_{i,l}]$ is a product of binomial distributions and $[\mathbf{v}_i|\mu,\sigma^2,\theta]$ is multivariate normal.  Then, given $\mathbf{v}_i$, the predictive full-conditional for $\tilde{\mathbf{v}}_i$ is 
\begin{equation}
  [\tilde{\mathbf{v}}_i|\mu,\sigma^2,\theta,\mathbf{v}_i]=\text{N}(\mu\mathbf{1}+\tilde{\boldsymbol\Sigma}\boldsymbol\Sigma^{-1}(\mathbf{v}_i-\mu\mathbf{1}),\tilde{\tilde{\boldsymbol\Sigma}}-\tilde{\boldsymbol\Sigma}\boldsymbol\Sigma^{-1}\tilde{\boldsymbol\Sigma}')\;,
  \label{eq:vpredfullcond}
\end{equation}
\noindent where $\tilde{\boldsymbol\Sigma}$ is the cross-covariance matrix between the predictions and observed data and $\tilde{\tilde{\boldsymbol\Sigma}}$ is the covariance matrix associated with the predictions.

To sample $\mathbf{v}_i$ from (\ref{eq:vfullcond}), there are a variety of approaches including standard MCMC with conditional Metropolis-Hastings updates, elliptical slice sampling \citep{murray2010elliptical}, Hamiltonian Monte Carlo \citep{neal1992bayesian}, and auxiliary variable approaches \citep{chib1998analysis}.  We used this latter approach in practice \citep{albert1993bayesian} to obtain Markov chains for $\mathbf{v}_i$ (via auxiliary variables) based on each parameter set $\mu^{(k)}$, $\sigma^{2(k)}$, $\theta^{(k)}$, for $k=1,\ldots,K$ (Appendix B).  Alternatively, this can be achieved with automated software (e.g., JAGS, NIMBLE, STAN; \citealt{plummer2003jags}, \citealt{NIMBLE}, \citealt{stan}) but may be slower and not result in well-mixed chains.  

To obtain inference for the individual-level utilization distributions we can use MC integration to compute 
\begin{equation}
  \frac{\sum_{k=1}^K \tilde{p}_i^{(k)}(\tilde{\mathbf{s}}_i)/ \tilde{\mathbf{p}}_i'^{(k)} \mathbf{1}}{K} \;,
  \label{eq:meanpostimplicitdistn}
\end{equation}
\noindent where $\tilde{\mathbf{p}}_i^{(k)}=\Phi(\tilde{\mathbf{v}}^{(k)})$ and $\tilde{p}_i^{(k)}(\tilde{\mathbf{s}}_i)$ is the $k$th posterior realization of detection probability at location $\tilde{\mathbf{s}}_i$.  For a fine grid of locations, this results in a map of individual level posterior space use.  However, the spatial pattern of (\ref{eq:meanpostimplicitdistn}) is the same as that of the posterior mean of $\tilde{\mathbf{p}}_i$ for individual $i$. Thus, we present the latter in the applications that follow.  

\section{Applications}
\subsection{Simulation}
\begin{figure}[htp]
  \centering
  \includegraphics[width=4in]{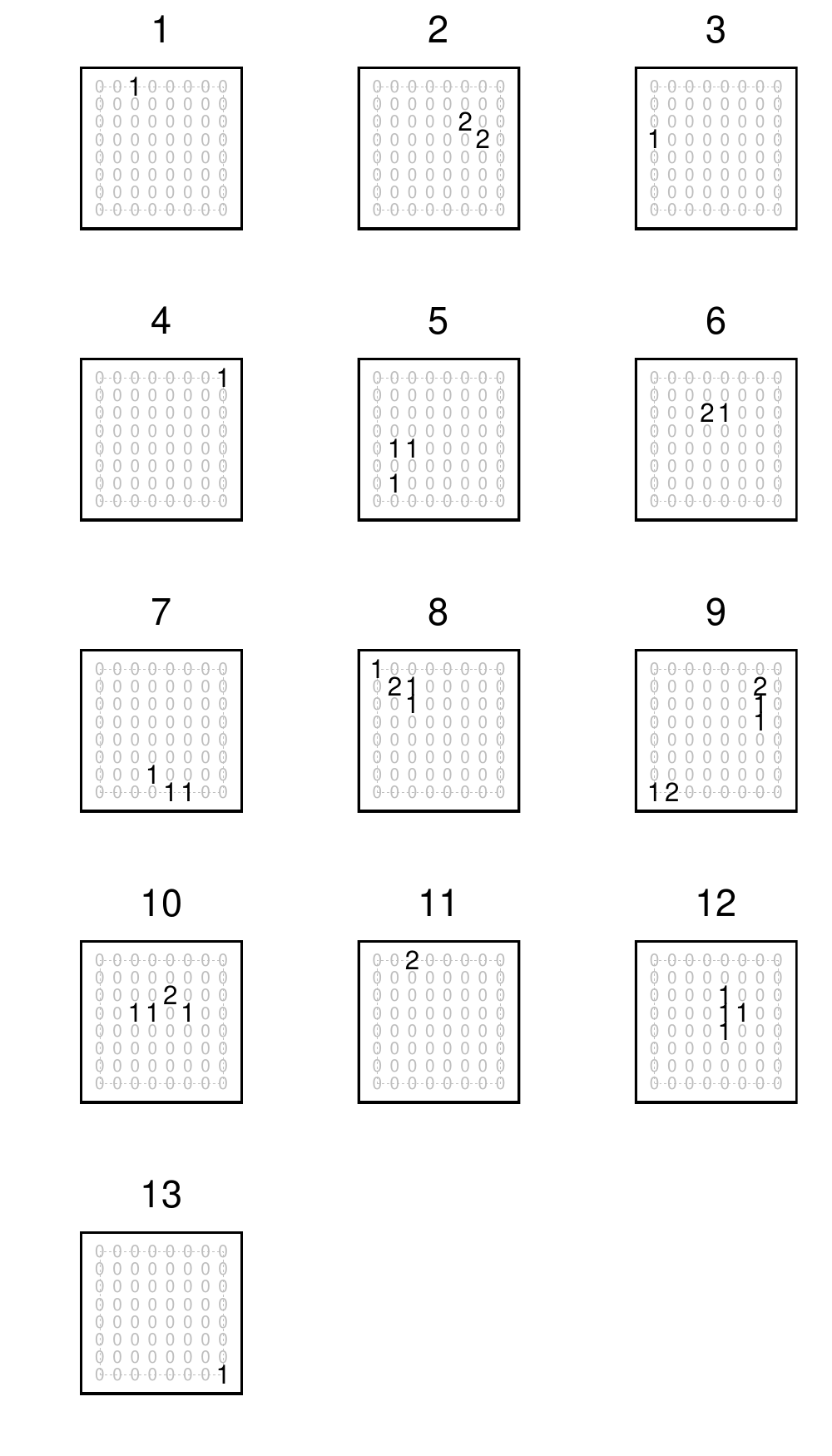}
  \caption{SCR data for $n=13$ simulated individuals over an $8\times 8$ array of $L=64$ traps spaced 0.143 units apart in a unit square study area.  Positions of numbers represent trap locations in the array and values correspond to the number of detections for each individual at each trap (cases with $y_{i,l}>0$ shown in bold large font).  Study area shown as a dashed box.}
  \label{fig:sim_data}
\end{figure}

We simulated capture-recapture data for a square trap array of $L=64$ equally spaced detectors on the unit square (Figure~\ref{fig:sim_data}).   
We used a multi-activity center SCR model to simulate the data.  This simulation model was based on a superpopulation of $M=200$ individuals, with $\psi=0.2$ membership probability which resulted in $N=29$ individuals in our simulated population. We simulated the number of activity centers for each individual from a zero-truncated Poisson distribution with intensity parameter equal to 0.5 and locations of the latent activity centers were sampled as a complete spatial random point process in a square buffered region centered on the unit square study area ${\cal S}$ containing our trap locations $\mathbf{x}_l$ for $l=1,\ldots,L$.  The buffered region extended $0.5$ units beyond the trap array in each direction to allow the activity centers for some individuals to occur outside the trap array boundary ${\cal S}$.  We used a maximum of the complementary log-log (`cloglog') link for the detection function from (\ref{eq:scr_link}) over all activity centers at each trap location with $\alpha=-1$ and $\beta=-50$.  These simulated data resulted in a mixture of unimodal and multimodal space use distributions for the $N$ individuals, with $n=13$ individuals observed (Figure~\ref{fig:sim_data}).  

We fit the GCR model as described in the Methods section with priors specified as $\mu\sim \text{N}(0,4)$, $\psi\sim \text{Beta}(1,1)$, and $\theta\sim \text{DiscUnif}(\Theta)$ with support on the set $\Theta\equiv\{\text{max}(\mathbf{D})/20,\ldots,\text{max}(\mathbf{D})/2\}$ of equally spaced values of $\theta$. The matrix $\mathbf{D}$ contains the pairwise distances among traps in our study area and $\text{max}(\mathbf{D})=\sqrt{2}$.  We set the superpopulation size to $M=200$ individuals for our study area and $\sigma^2=1$.  

We used the multistage computing strategy to fit the GCR model with $K=100000$ MCMC iterations.  Stage one required 61.4 minutes and stage two required 1.2 minutes.  After fitting the model, \emph{post hoc} sampling of the abundance $N$ required 5.4 minutes and then we used a Gibbs sampler to obtain an MCMC sample for $\mathbf{P}$ for the simulated individuals 9 and 12.  A Gibbs sampler required approximately 36 seconds for each individual.  All computation was performed on a machine with a 24-core 3.6 Ghz processor and 192 MB of RAM. 

\begin{figure}[htp]
  \centering
  \includegraphics[width=5in]{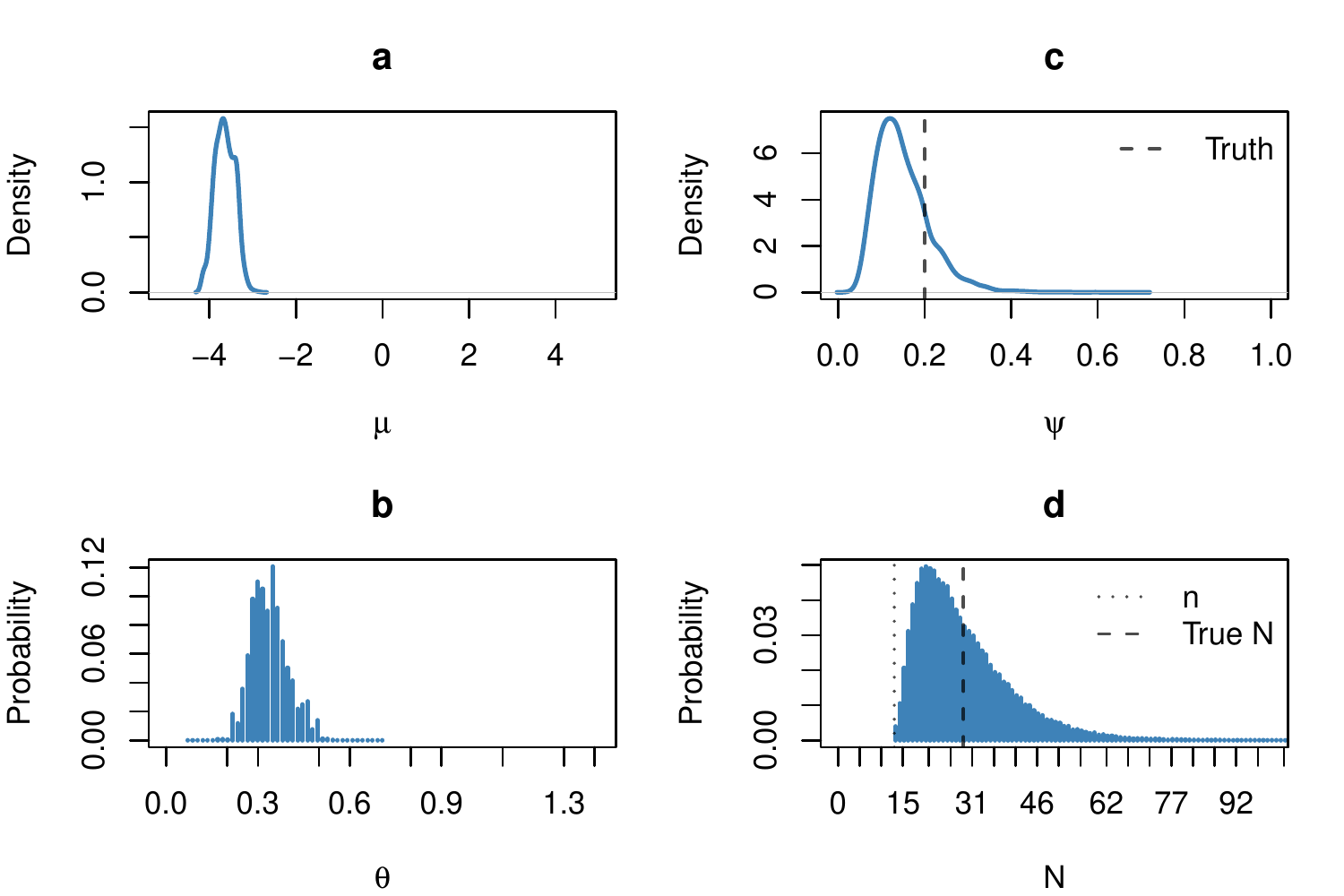}
  \caption{Marginal posterior distributions for GCR parameters $\mu$ (panel a), $\theta$ (panel b), $\psi$ (panel c), and $N$ (panel d) resulting from our simulated data.  Panel d shows both the observed number of individuals $n$ (dotted vertical) and true number of individuals $N$ (dashed vertical).}
  \label{fig:sim_params_post}
\end{figure}

The GCR model fitted to the simulated data resulted in the marginal posterior distributions shown in Figure~\ref{fig:sim_params_post}.  
Figure~\ref{fig:sim_params_post}a shows a small estimated $\mu$ which implies that the space use throughout most of the study area is low for a given individual.  If the species occupied more of the study area, the parameter $\mu$ could increase to accommodate that type of space use.  

The marginal posterior mean for population abundance $N$ was $28.5$ (compared with the true simulated value $N=29$; Figure~\ref{fig:sim_params_post}d).  The 95\% marginal posterior credible interval for abundance $N$ was $(15,55)$ and the marginal posterior quartiles were 21 and 34.   

\begin{figure}[htp]
  \centering
  \includegraphics[width=5in]{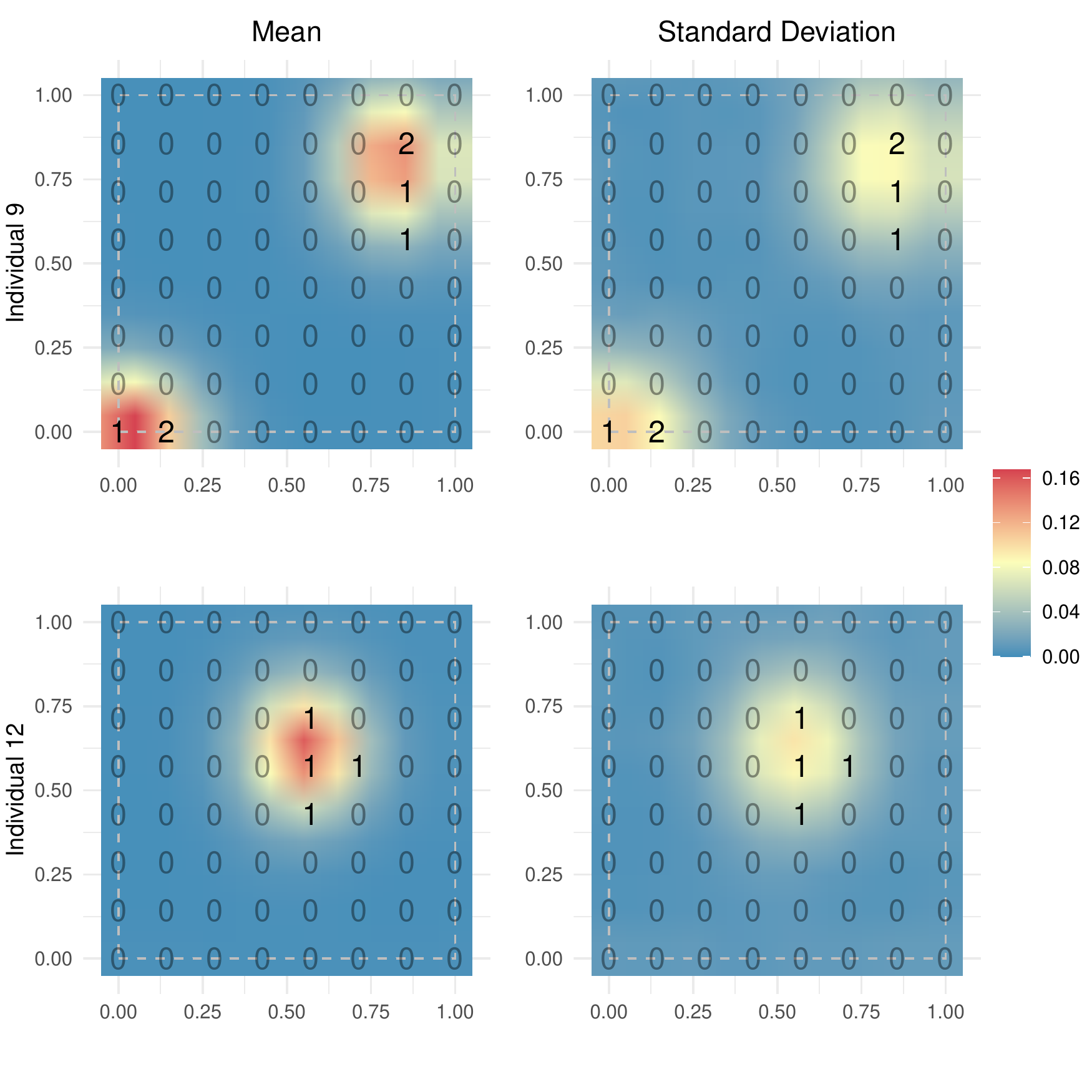}
  \caption{Posterior mean and standard deviation of $\tilde{p}$ for simulated individuals 9 and 12 in our study area.  Count of detections for each trap at $\mathbf{x}_l$ for $l=1,\ldots,64$ shown as numbers.}
  \label{fig:sim_p_post}
\end{figure}

Figure~\ref{fig:sim_p_post} shows the posterior mean (and posterior standard deviation) for the implied space use distribution for two observed individuals in our simulated data; individual 9 illustrates a bimodal space use pattern and individual 12 illustrates a unimodal space use pattern.   
For comparison, we also fit the traditional SCR model with detection function in (\ref{eq:scr_datamodel}) to the same simulated data (Appendix C), but with the buffer included  in the state-space for the latent activity center point process.  The results indicate a unimodal symmetric space use pattern between the two sets of detections for individual 9 (Figure~\ref{fig:sim_scr_p_post}) rather than the bimodal space use pattern resulting from a fit of the GCR model.   

\subsection{Snowshoe Hare}
\begin{figure}[htp]
  \centering
  \includegraphics[width=4in]{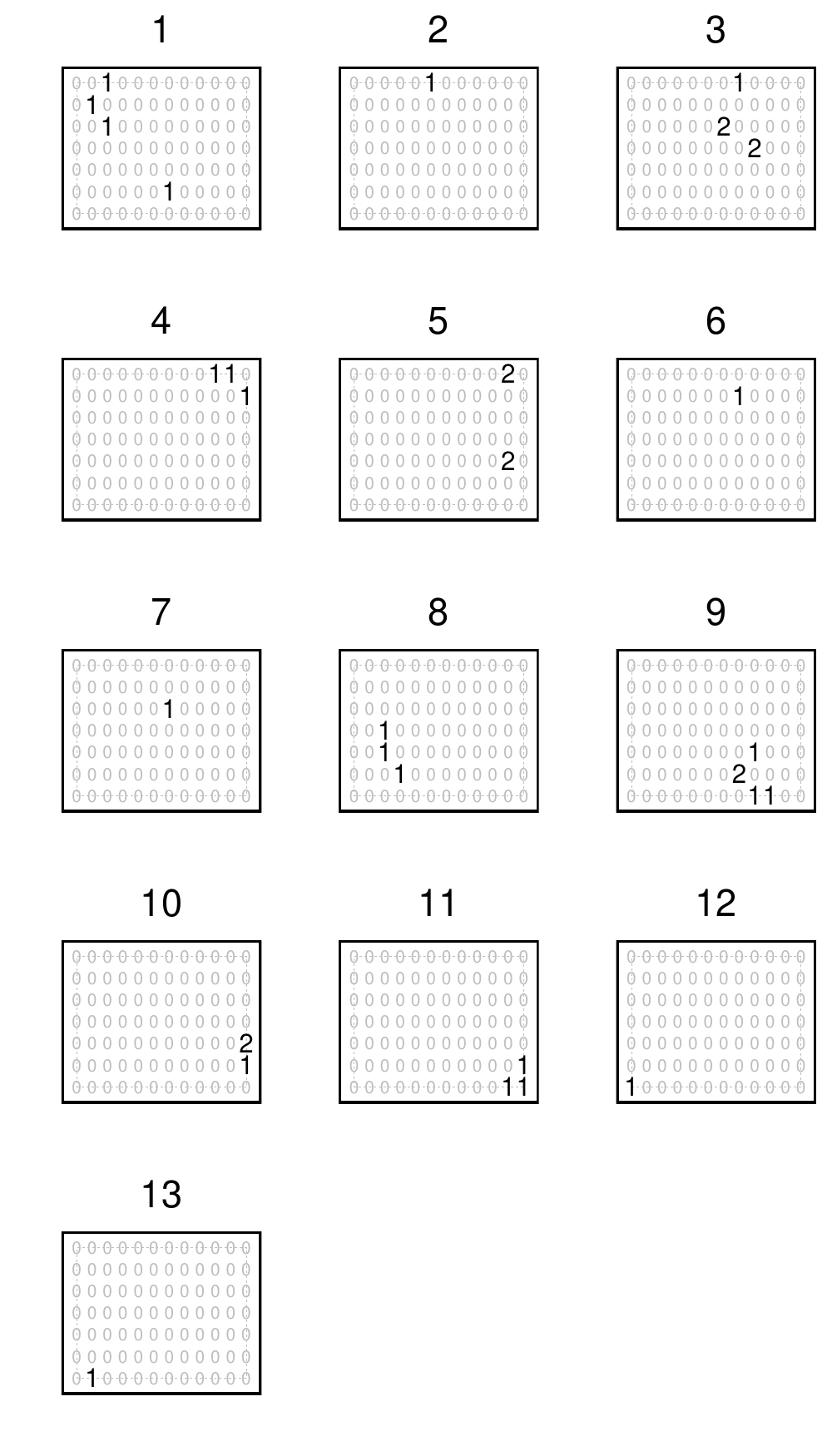}
  \caption{SCR data for $n=13$ snowshoe hare individuals over a $7\times 12$ array of $L=84$ traps spaced 50m apart.  Positions of numbers represent trap locations in array and values correspond to the number of detections for each individual at each trap (cases with $y_{i,l}>0$ shown in bold).  Study area is shown as a dashed box.}
  \label{fig:SS_data}
\end{figure}

We applied the GCR model to analyze a set of spatially-explicit capture histories of $n=13$ snowshoe hares in Colorado, USA.  The data contain counts of detections over $J=5$ sampling occasions at $L=84$ trap locations (Figure~\ref{fig:SS_data}) and were collected during winter 2007 \citep{ivan2014density}.  We defined the study area as the convex hull of the trap locations for this analysis.  Exploratory analysis of these data suggest that certain individuals may not use space in a way that meets the assumptions of a standard SCR model because their detections occurred farther apart than those for most other individuals in the study; for example, individuals 1 and 5 in Figure~\ref{fig:SS_data}.   

We fit the GCR model as described in the Methods section with priors specified as $\mu\sim \text{N}(0,4)$, $\psi\sim \text{Beta}(1,1)$, and $\theta\sim \text{DiscUnif}(\Theta)$ with support on the set $\Theta\equiv\{\text{max}(\mathbf{D})/20,\ldots,\text{max}(\mathbf{D})/2\}$ of equally spaced values, where $\mathbf{D}$ is the pairwise distance matrix among traps in our study area.  We set the superpopulation size to $M=200$ individuals for our study area and $\sigma^2=1$ to ensure the other parameters were identifiable. 

We used $K=100000$ MCMC iterations to fit the GCR model.  This required 73.3 minutes for stage one, and only 1.4 minutes for stage two of the recursive procedure.  To sample the abundance $N$ required 7.2 minutes and to sample from the full-conditional for $\mathbf{v}$ for an individual required approximately 54 seconds using a Gibbs sampler.  All computation was performed on a machine with a 24-core 3.6 Ghz processor and 192 MB of RAM. 

\begin{figure}[htp]
  \centering
  \includegraphics[width=5in]{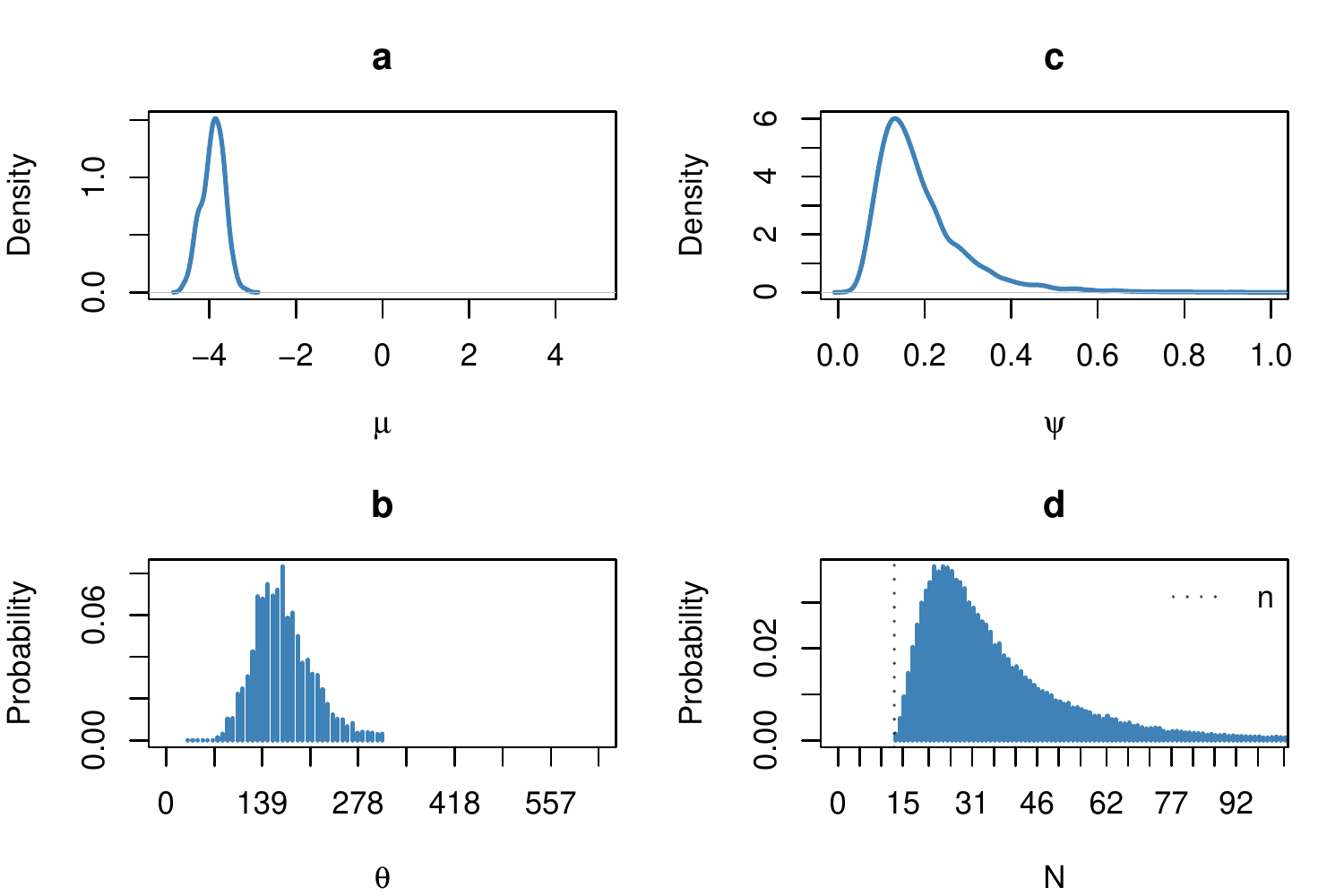}
  \caption{Marginal posterior distributions for GCR parameters $\mu$ (panel a), $\theta$ (panel b), $\psi$ (panel c), and $N$ (panel d).  Panel d shows the observed number of individuals $n$ (dotted vertical).}
  \label{fig:SS_params_post}
\end{figure}

The GCR model fitted to the snowshoe hare data resulted in marginal posterior distributions shown in Figure~\ref{fig:SS_params_post}.  
Figure~\ref{fig:SS_params_post} indicates that a small $\mu$ and large $\theta$ were needed to result in adequate smoothness of the space use distribution throughout the study area.  Furthermore, the marginal posterior 95\% credible interval for abundance $N$ was $(16,88)$ and the marginal posterior quartiles were 24 and 43.    

\begin{figure}[htp]
  \centering
  \includegraphics[width=5in]{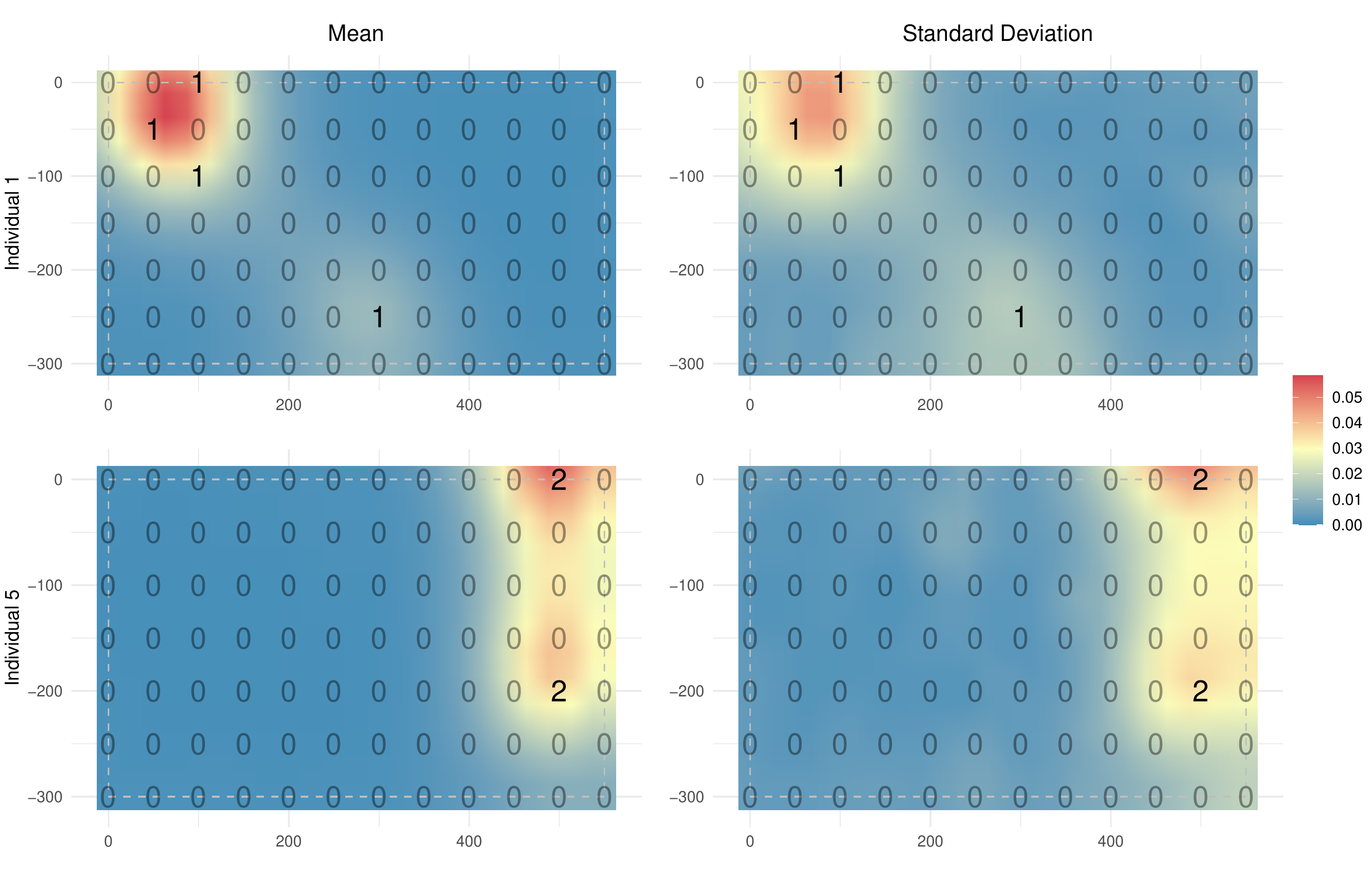}
  \caption{Posterior mean and standard deviation of $\tilde{p}$ for snowshoe hare individuals 1 and 5 in our study area (Figure~\ref{fig:SS_data}).  Count of detections for each trap at $\mathbf{x}_l$ for $l=1,\ldots,84$ shown as numbers.}
  \label{fig:SS_p_post}
\end{figure}

For example, Figure~\ref{fig:SS_p_post} shows the posterior mean (and posterior standard deviation) for the implied space use distribution for two observed individuals in our study that have clearly bimodal space use patterns.   
These types of individual-level detection probability patterns (and hence space use patterns) (Figure~\ref{fig:SS_p_post}, left column) would not be possible under the conventional SCR model.  

\section{Discussion}
We presented a formulation of spatial capture-recapture models that allows for structured heterogeneity in detection probability over the study area.  Our approach assumes the detection probability surface (and hence implied space use distribution) can be characterized by a GP.  We refer to this new CR model formulation as ``geostatistical'' to associate it with the previous studies that have used GPs for modeling other ecological and environmental processes.  While there are a variety of approaches to fitting such models to data, we showed how to leverage multicore computing resources based on a multistage computing strategy.  By extending the procedure described by \cite{hooten2023multistagecr} to accommodate latent GPs, we showed that the GCR model was able to recover total abundance and detection probability patterns that are not unimodal using simulated data.  

We also fit the GCR model to capture-recapture data on snowshoe hares in Colorado, USA.  Some of the capture histories associated with our observed snowshoe hares showed evidence of multimodality and asymmetric detection probability patterns.  The posterior predictive space use distributions we obtained by fitting our GCR model confirmed this.  

Whereas conventional SCR models are specified based on an assumption about latent point processes, our approach relies on the flexibility of latent continuous spatial processes.  We appreciate the mechanism involved in the conventional SCR model specifications and our GCR model lacks that direct connection to an explicit activity center for each individual.  However, we also see the need for additional flexibility in some cases where the space use distributions may not adhere to single elliptical shapes.  Furthermore, the latent GP is not without mechanistic connections.  For example, in ongoing work, we are exploring the use of multi-output GPs \citep{alvarez2012kernels} in GCR models that can account for dependence among individuals explicitly.  For example, territorial species may use space that is unoccupied by conspecifics.  Therefore we expect those individual space use distributions to be negatively correlated with each other.  Our GCR model allows for both negative and positive interactions among individuals to be accounted for using the same types of spatial statistical approaches that have been used to account for dependence among other multivariate spatial processes in nature \citep[e.g., coregionalization and cokriging approaches;][]{journel1978mining,higdon2002space}.

For those interested in multiscale ecological processes, a simpler extension of the GCR approach could facilitate separate inference for individual-based space use and species distribution by allowing for a shared spatially heterogeneous mean function in the latent GP across individuals.  That mean function would then represent species distribution due to environmental factors that relate to the fundamental niche.  Then the individual-specific spatial random field could account for resource selection and movement at the finer scale.   

In fact, a rapidly growing area of development in SCR modeling involves the use of telemetry data \citep{hooten2017animal} coupled with CR data to account for individual-level movement patterns in abundance estimation \citep[e.g.,][]{mcclintock2022integrated}.  Our GCR model could be paired with a telemetry data model in a similar way when both sources of data exist.  

\section*{Acknowledgments} 
This research was funded by NSF 1614392 and NSF 2222525.  The authors thank the editors and anonymous reviewers of Spatial Statistics as well as Matthias Katzfuss and Antik Chakraborty for helpful discussions and insights about this research.   

\spacingset{1} 
\bibliographystyle{ecology} 
\bibliography{cr_refs,scr_refs}

\clearpage

\spacingset{1.45} 
\section*{Appendix A}
To show how the form of model that involves conditioning on observed sample size $n$ in (\ref{eq:full_post}) arises, we demonstrate with a simple capture-recapture (CR) model for brevity in what follows.  Extensions to the case where detection probability varies are described in \citet{hooten2023multistagecr}.  

Suppose we have a CR model with $y_i$ modeled as in (\ref{eq:scr_datamodel}) but with homogeneous detection probability $p$ and population membership indicator $z_i \sim \text{Bern}(\psi)$ for $i=1,\ldots,M$.  In this case, using PX-DA, $y_i=0$ for undetected individuals $i=n+1,\ldots,M$.  This results in two sets of data:  The original observed data $\mathbf{y}_{1:n}\equiv (y_1,\ldots,y_n)'$ and the augmented zero data $\mathbf{y}_{(n+1):M}\equiv(y_{n+1},\ldots,y_M)'$.  If we marginalize over $z_i$ to yield the ``integrated'' data model $[y_i|p,\psi]$, then we can write the posterior distribution of interest as
\begin{equation}
  [p,\psi|\mathbf{y}_{1:n},\mathbf{y}_{(n+1):M},n] = [\mathbf{y}_{(n+1):M}|p,\psi,\mathbf{y}_{1:n},n][p,\psi|\mathbf{y}_{1:n},n] \;, 
\end{equation}
where we condition on $n$ as an auxiliary source of data because it is unknown until observed as part of the CR survey, like $\mathbf{y}_{1:n}$.  The first term on the right-hand side in the above expression is irrelevant because the elements of $\mathbf{y}_{(n+1):M}$ are always known to equal zero when $n$ is observed.  Thus, the second term in the above expression can be factored as 
\begin{equation}
  [p,\psi|\mathbf{y}_{1:n},n] \propto \left(\prod_{i=1}^n [y_i|p,y_i>0] \right) [n|p,\psi][p][\psi] \;,
\end{equation}
where the conditional distribution $[y_i|p,y_i>0]$ is a zero-truncated binomial and the conditional distribution $[n|p,\psi]$ is either Poisson-binomial or Poisson depending on model assumptions because $n$ is the sum of binary variables indicating which individuals in the superpopulation were detected.  In our GCR model, the detection probability is heterogeneous and stochastic, thus the conditional distribution for $n$ is obtained by marginalizing over $p$.  

In terms of implementation of this simple model, for the first stage, we can use an MCMC algorithm with Metropolis-Hastings updates based on a proposal for $p$ from its prior (or a random walk).  The membership probability $\psi$ can be Monte Carlo sampled directly from its prior because it does not appear in the first-stage conditional data model.  In the GCR model, we only have two parameters $\mu$ and $\theta$.  We used a Gaussian random-walk proposal for $\mu$ and a discrete random-walk proposal for $\theta$ and then a block update for both parameters simultaneously.   

\section*{Appendix B}
To sample $\mathbf{v}_i$ (and hence $\mathbf{p}_i$) from its full-conditional distribution after implementing computing stages 1 and 2, we used a Gibbs sampler following a reparameterization of the distribution to include auxiliary variables \citep{albert1993bayesian}.  This approach was stable and fast relative to sampling using automated software.

As a brief summary of the approach, to sample from the full-conditional distribution of $\mathbf{v}_i$, we recognize that we simply need to use MCMC to fit the model $y_{i,l} \sim \text{Binom}(J,p_{i,l})$ where $\mathbf{v}_i=\Phi^{-1}(\textbf{p}_i)\sim \text{N}(\mu \mathbf{1}, \mathbf{R}(\theta))$ assuming that $\mu$ and $\theta$ are known (from the stage 2 MCMC output).  It is not necessary to sample $\mathbf{v}_i$ to fit the broader GCR model because it has been marginalized out in stages 1 and 2, but if we desire inference about the detection probability surface for certain individuals (e.g., Figures~\ref{fig:sim_p_post} and \ref{fig:SS_p_post}), we can obtain a sample for $\mathbf{v}_i$ \emph{post hoc}.  We can do this by fitting the model described above while conditioning on each realization of $\mu^{(k)}$ and $\theta^{(k)}$ from the stage 2 output and retain a single realization of $\mathbf{v}^{(k)}_i$ (e.g., the last realization in a short chain after burn-in) for each $k=1,\ldots,K$.   

For a specific individual $i$ and conditional on $\mu$ and $\theta$, we need only fit a Bayesian binomial generalized linear model (GLM) with probit link function $\Phi^{-1}(\textbf{p}_i)=\mu\mathbf{1} + \boldsymbol\eta_i$ with prior $\boldsymbol\eta_i \sim \text{N}(\mathbf{0},\mathbf{R}(\theta))$.  Thus, the random vector $\boldsymbol\eta_i$ can be treated as a set of parameters in this reduced model. Alternatively, we can write $\boldsymbol\eta_i=\mathbf{H}\boldsymbol\alpha_i$ where $\boldsymbol\alpha_i \sim \text{N}(\mathbf{0},\boldsymbol\Lambda)$ and $\mathbf{H}$ is a matrix of spatial basis functions (e.g., we used eigenvectors resulting from the spectral decomposition $\mathbf{R}(\theta)=\mathbf{H}\boldsymbol\Lambda\mathbf{H}'$ and $\boldsymbol\Lambda$ is a diagonal matrix of associated eigenvalues).  This latter linear model formulation is not strictly necessary but could be used to reduce dimensionality to improve computation time by truncating the matrix $\mathbf{H}$ and associated dimension of the spectral coefficients $\boldsymbol\alpha_i$. 
Based on the probit link function and multivariate normal prior for $\boldsymbol\alpha_i$, we can write the model for a given individual of interest $i$ as: 
\begin{equation}
  y_{i,l,j} = 
  \begin{cases}
    0, z_{i,l,j} \leq 0 \\
    1, z_{i,l,j} > 0 
  \end{cases} \;,
\end{equation}
\noindent where $z_{i,l,j} \sim \text{N}(\mu + \mathbf{h}'_l\boldsymbol\alpha_i, 1)$ with prior $\boldsymbol\alpha_i \sim \text{N}(\mathbf{0},\boldsymbol\Lambda)$ for $j=1,\ldots,J$ sampling occasions and $\mathbf{h}'_l$ is the $l$th row vector of $\mathbf{H}$.  Jointly for an individual of interest $i$, we can write the model for the $L\times 1$ latent random vector $\mathbf{z}_{i,j}$ as   
\begin{equation}
  \mathbf{z}_{i,j} \sim \text{N}(\mu \mathbf{1}+\mathbf{H}\boldsymbol\alpha_i, \mathbf{I}) \;.
\end{equation}
This results in the following tractable full-conditional distributions  
\begin{align}
  [z_{i,l,j}|\cdot] &= 
  \begin{cases}
     \text{TN}(\mu + \mathbf{h}'_l\boldsymbol\alpha_i,1)^0_{-\infty} &\mbox{, } y_{i,j,l}=0 \\
     \text{TN}(\mu + \mathbf{h}'_l\boldsymbol\alpha_i,1)^{\infty}_0 &\mbox{, } y_{i,j,l}=1 \\
  \end{cases} \;, \\
  [\boldsymbol\alpha_i|\cdot] &= \text{N}((J\mathbf{H}'\mathbf{H}\boldsymbol+\boldsymbol\Lambda)^{-1}(\mathbf{1}\otimes\mathbf{H})'(\mathbf{z}_i-\mu\mathbf{1}),(J\mathbf{H}'\mathbf{H}\boldsymbol+\boldsymbol\Lambda)^{-1}) \;, 
\end{align}
where $\mathbf{z}_i=(z_{i,1,1},z_{i,2,1},\ldots,z_{i,l,j},\ldots,z_{i,L-1,J},z_{i,L,J})'$. These full-conditional distributions can be sampled from sequentially in our secondary MCMC algorithm and we obtain the realization $\mathbf{v}^{(k)}_i = \mu\mathbf{1}+\mathbf{H}\boldsymbol\alpha^{(k)}_i$ as a derived quantity.  To predict $\tilde{\mathbf{v}}_i$ at a grid of non-trap locations throughout the study area, we sample from its predictive full-conditional distribution (\ref{eq:vpredfullcond}) as described in the Inference and Prediction section, then transform to $\tilde{\mathbf{p}}^{(k)}_i$ and use Monte Carlo integration to obtain posterior moments (i.e., the maps in Figures~\ref{fig:sim_p_post} and \ref{fig:SS_p_post}).

\section*{Appendix C}
To compare inferred space use patterns, we fit the traditional SCR model from (\ref{eq:scr_datamodel}) to the simulated data shown in Figure~\ref{fig:sim_data} using JAGS.  We used the spatial domain from the simulation (extended $0.5$ units beyond the trap array to account for activity center that may fall outside the trap array) and specified vague Gaussian priors for the SCR parameters $\alpha$ and $\beta$, and a uniform prior for $\psi$.  We also used the same superpopulation size $M=200$.  
 
The GCR model is able to account for irregular detection probability patterns in space, but the SCR model is not.  The results of fitting the classical SCR model to our data can be compared with those from the GCR model in terms of the detection probability $\tilde{p}$ for a grid of prediction locations.  
\begin{figure}[htp]
  \centering
  \includegraphics[width=5in]{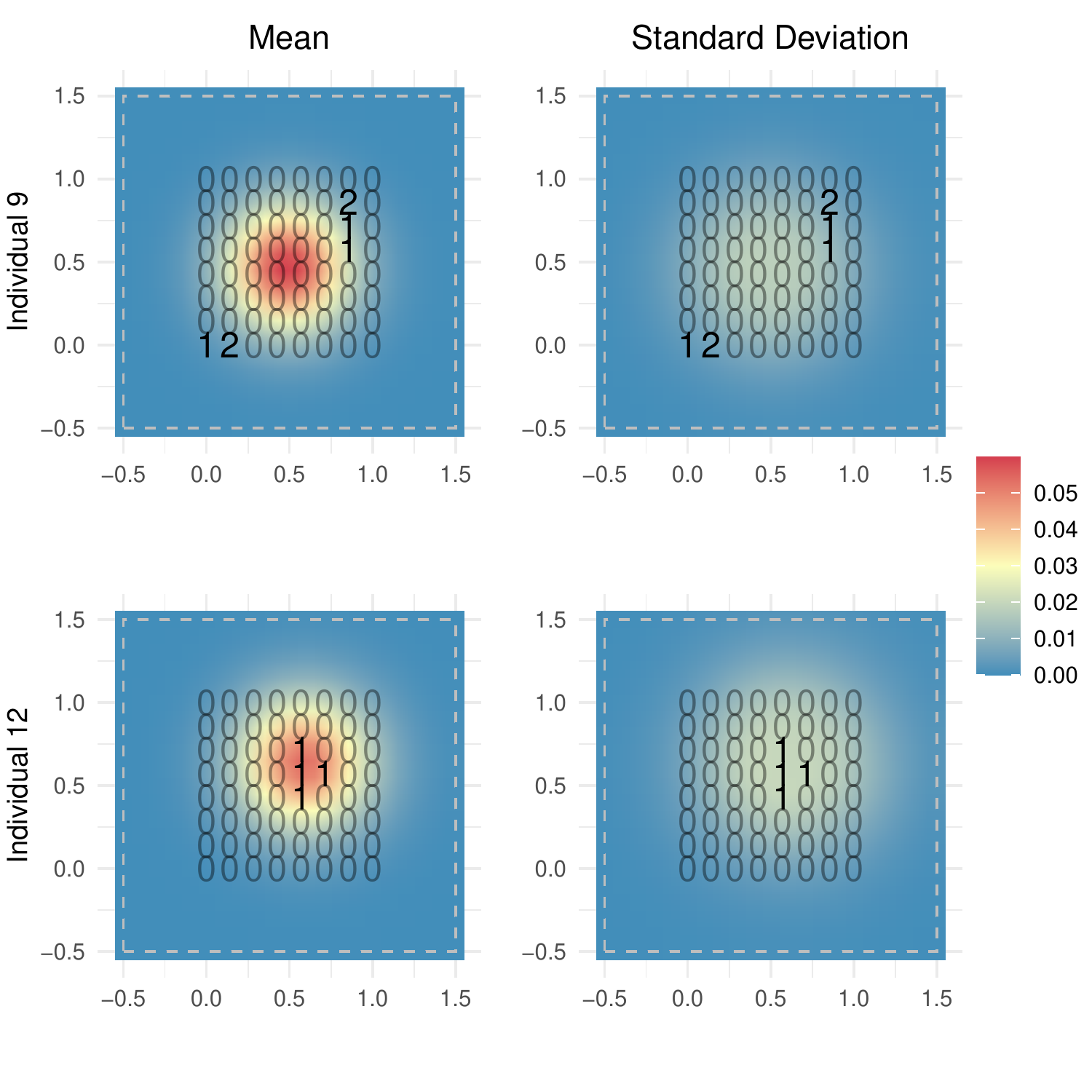}
  \caption{Posterior mean and standard deviation of $\tilde{p}$ from the traditional SCR model for simulated individuals 9 and 12 in our study area.  Count of detections for each trap at $\mathbf{x}_l$ for $l=1,\ldots,64$ shown as numbers.  The sampled region is buffered to allow for individuals with activity centers outside the trap array.}
  \label{fig:sim_scr_p_post}
\end{figure}

We note that the predicted detection probability function places the center of probability mass between the two clusters of detections for individual 9 (Figure~\ref{fig:sim_scr_p_post}).  The radial shape evident in the individual space use patterns is a characteristic of the SCR model and allows it to be powerful when the underlying detection probability meets the assumptions (e.g., for individual 12), but as in this case with simulated individual 9, it may not appropriately represent more complicated patterns (as compared with the GCR results presented in Figure~\ref{fig:sim_p_post}).  From a model checking perspective \citep[e.g.,][]{conn2018guide}, we could omit the classical SCR model from our set of candidate models based on these results.  

\end{document}